\documentclass[a4paper,fleqn]{cas-sc}

\usepackage[square,numbers]{natbib}
\usepackage{graphicx}
\usepackage[normalem]{ulem}
\useunder{\uline}{\ul}{}
\usepackage{breakurl}
\usepackage{tabularx}
\usepackage{booktabs}
\usepackage{eurosym}
\usepackage{breakurl}
\usepackage{todonotes}
\usepackage[normalem]{ulem}
\usepackage{multirow}
\usepackage{tabulary}
\usepackage{hyperref}
\usepackage{color}
\usepackage{amssymb}
\usepackage{rotating}
\usepackage[T1]{fontenc}
\usepackage{lscape}
\usepackage[autostyle]{csquotes}
\usepackage[inline]{enumitem}
\usepackage[ruled,vlined]{algorithm2e}

\usepackage{siunitx}
\usepackage{longtable}
\usepackage[utf8]{inputenc}
\usepackage{longtable}
\usepackage{pdflscape}
\usepackage{graphicx}
\usepackage{tabularx}
\usepackage{multirow}
\usepackage{textcomp}
\usepackage{csquotes}
\usepackage{booktabs}
\usepackage{graphics}
\usepackage{xr-hyper}
\usepackage{hyperref}
\usepackage{amsmath}
\usepackage{amssymb}
\usepackage{natbib}
\usepackage{epsfig}
\usepackage{lscape}
\usepackage{lineno}
\usepackage{array}
\usepackage{float}
\usepackage{todonotes}
\usepackage{tabu}
\usepackage{siunitx}

\usepackage{pdflscape}
\usepackage{setspace}
\usepackage{textcomp}
\definecolor{bblue}{HTML}{4285F4}
\definecolor{rred}{HTML}{DB4437}
\definecolor{ggreen}{HTML}{0F9D58}
\definecolor{yyellow}{HTML}{F4B400}
\usepackage{pgfplots}
\usepgfplotslibrary{units}
\pgfplotsset{compat=1.14}

\def\tsc#1{\csdef{#1}{\textsc{\lowercase{#1}}\xspace}}
\tsc{WGM}
\tsc{QE}
\tsc{EP}
\tsc{PMS}
\tsc{BEC}
\tsc{DE}

\begin{document}
\let\WriteBookmarks\relax
\def\floatpagepagefraction{1}
\def\textpagefraction{.001}

\shorttitle{Unveiling Hidden Energy Anomalies}

\shortauthors{F. Fadli et~al.}

\title [mode = title]{Unveiling Hidden Energy Anomalies: Harnessing Deep Learning to Optimize Energy Management in Sports Facilities}                      

\vskip2mm

\author[1]{Fodil Fadli\corref{cor1}}
\ead{f.fadli@qu.edu.qa}
\author[2]{Yassine Himeur}
\ead{yhimeur@ud.ac.ae}
\author[3]{Mariam Elnour}
\ead{mariam.elnour@outlook.com}
\author[4,5]{Abbes Amira}
\ead{aamira@sharjah.ac.ae}

\address[1]{Department of Architecture \& Urban Planning, Qatar University, Doha, Qatar}
\address[2]{College of Engineering and Information Technology, University of Dubai, Dubai, UAE}
\address[3]{Department of Electrical Engineering, Qatar University, Doha, Qatar}
\address[4]{Department of Computer Science, University of Sharjah, UAE}
\address[5]{Institute of Artificial Intelligence, De Montfort University, Leicester, United Kingdom}


%
%
%
%
%
%
%
%
%
%
%

\begin{abstract}
Anomaly detection in sport facilities has gained significant attention due to its potential to promote energy saving and optimizing operational efficiency. In this research article, we investigate the role of machine learning, particularly deep learning, in anomaly detection for sport facilities. We explore the challenges and perspectives of utilizing deep learning methods for this task, aiming to address the drawbacks and limitations of conventional approaches.
Our proposed approach involves feature extraction from the data collected in sport facilities. We present a problem formulation using Deep Feedforward Neural Networks (DFNN) and introduce threshold estimation techniques to identify anomalies effectively. Furthermore, we propose methods to reduce false alarms, ensuring the reliability and accuracy of anomaly detection.
To evaluate the effectiveness of our approach, we conduct experiments on aquatic center dataset at Qatar University. The results demonstrate the superiority of our deep learning-based method over conventional techniques, highlighting its potential in real-world applications. Typically, 94.33\% accuracy and 92.92\% F1-score have been achieved using the proposed scheme. 
\end{abstract}




\maketitle

\section{Introduction} 
\subsection{Preliminary}
According to recent statistics, current energy policies project a substantial 84\% increase in global electricity consumption over the next 25 years. This surge in demand makes energy efficiency a crucial element for countries worldwide \cite{lei2023dynamic,al2022interactive}. For instance, the EU has established ambitious objectives for energy and climate policy, exemplified by the 2030 goals (aiming at least a 55\% reduction in greenhouse gas emissions from 1990 levels, significantly upping the ambition from the original 40\% target, at least a 32.5\% improvement in energy efficiency compared to projections, and at least a 40\% share of renewable energy sources in the EU's energy mix.). Additionally, the EU is striving for even more significant achievements with the 40/27/27 goals, targeting a 40\% domestic reduction in greenhouse gas emissions, 27\% improvement in energy efficiency, and 27\% renewable energy by 2030 \cite{sayed2022artificial,shekhar2023demand}.
To attain these goals, information and communication technologies (ICT) can play a pivotal role in driving energy efficiency. However, despite the potential benefits, many consumers and managers remain hesitant to adopt these technologies due to several reasons \cite{elnour2022neural}. Firstly, there is a lack of sufficient demonstrations showcasing the cost-effectiveness of ICT solutions \cite{himeur2022next}. Secondly, the current under-development of applications that harness energy usage data for the benefit of consumers and third parties creates uncertainty \cite{varlamis2022using}. Thirdly, consumers are concerned about achieving energy savings without compromising comfort levels. Finally, limitations in consumers' capacity or capability to adopt energy-efficient practices pose additional challenges \cite{kong2023anomaly,alsalemi2022innovative}.
Moreover, most building customers are not adequately exposed to price signals provided through diverse pricing models and rate designs, which can further hinder their motivation to reduce energy consumption \cite{lin2023anomaly,varlamis2022smart}.
In light of these challenges and goals, it becomes imperative for policymakers, businesses, and individuals to collaborate and implement innovative solutions that promote energy efficiency while addressing consumer concerns \cite{wang2023real,sayed2021intelligent}. By leveraging the potential of ICT, providing effective demonstrations of cost-effectiveness, and enabling informed energy consumption decisions through pricing models, we can move closer to a sustainable and energy-efficient future \cite{lazim2023embedded,bousbiat2023neural}.

The upward trend in building energy consumption is a significant concern worldwide, driven by ongoing technological advancements, and urbanization, population growth, and rising living standards \cite{himeur2023ai}. As a consequence, buildings account for a substantial portion of global energy consumption, contributing to nearly 40\% of it, while also being responsible for up to 45\% of CO$_{2}$ emissions \cite{benavente2020buildings,homod2020evaluation}. Despite recent efforts to explore renewable energy and green building approaches, these solutions often remain cost-prohibitive and inaccessible to a large segment of the population \cite{himeur2022recent}.
In response to these challenges, there is a growing interest in adopting innovative technologies such as ICT, artificial intelligence (AI), machine learning (ML), internet of things (IoT), edge and cloud computing, to drive energy savings in buildings \cite{manimala2021role}. The integration of these technologies, alongside advanced metering infrastructure, smart distribution boards, and renewable energy sources, has led to the development of the concept of a "smart grid (SG) \cite{bousbiat2023crossing}."

As per \cite{doe2015assessment}, the rise in energy consumption in buildings can be attributed to several factors. One significant factor is that many buildings do not operate as intended by their designers and managers. Incorrect energy consumption habits of occupants, lack of awareness, malfunctioning equipment, and faulty devices, along with improper operating procedures and wrongly configured monitoring systems collectively contribute to buildings consuming approximately 20\% more energy than necessary \cite{perri2020smart}. Moreover, building systems may fail to meet performance expectations due to various faults, such as poorly maintained, degraded, and improperly controlled equipment, resulting in an estimated wastage of 15\% to 30\% of energy used in commercial buildings \cite{li2021impact,li2023behavior}. To address this issue, it becomes imperative to develop automated, swift, precise, and dependable fault detection and diagnosis schemes to ensure optimal system operations and achieve energy conservation \cite{yang2023innovative}.

Efficient building energy-saving systems necessitate early identification of unusual energy consumption and the prevention of energy frauds in smart meters. These systems enable operators and end-users to anticipate uncommon events, recognize abnormal energy consumption behaviors, and detect irregular energy usage. Anomaly detection of energy consumption has emerged as a highly sought-after area of research, drawing considerable attention from both the artificial intelligence and energy research communities \cite{himeur2021artificial}. Energy consumption anomalies can take the form of pattern anomalies or contextual anomalies. Pattern anomalies typically appear as outliers, with values significantly deviating from neighboring energy consumption data. They may arise from device or system malfunctions, noise impulses when devices are turned on, or temporary interference in sensor readings. On the other hand, contextual anomalies represent sets of energy consumption patterns that deviate from the norm within a specific period, even though their individual values fall within the normal energy consumption range. This type of anomaly can be caused by factors such as excessive energy usage due to end-users' behavior, faulty devices or systems consuming more energy than expected, or energy-hungry devices \cite{liu2020anomaly,himeur2021smart}.

Anomaly detection is the act of pinpointing events or instances that deviate from the usual patterns or norms. These unusual occurrences can shed light on a system's well-being, highlighting issues or potentially harmful activities. With the rise of IoT and intelligent technologies that produce massive data streams, data-centric and notably machine learning methods have grown in popularity for detecting these anomalies. Machine learning techniques for anomaly detection are now prevalent in numerous areas such as fault detection in heating, ventilation, and air conditioning (HVAC) systems \cite{elnour2020sensor,yun2021data, han2019least, bang2019novel}, security breaches in industrial controls \cite{noorizadeh2021cyber, elnour2020dual, karimipour2019deep, elnour2021application}, spotting fraud \cite{rtayli2020enhanced, paruchuri2017credit}, and identifying unusual energy consumption in structures \cite{chiosa2021data, rashid2018monitor, sial2019detecting}. Using machine learning for anomaly detection simplifies the process, negating the need for specialist input or handcrafted rules. A high-quality dataset and meticulous calibration of the ML technique are primarily what's required. Although calibration can be extensive, modern technology advancements facilitate its automation. Furthermore, employing machine learning in this domain often leads to enhanced results, as optimized ML models tend to outperform conventional methods in terms of accuracy and reliability.

In recent years, deep learning has made impressive strides in tackling problems of large data dimensions, interrelated data, and variability across numerous applications \cite{pang2021deep}. However, traditional deep learning methods were not fit for anomaly detection because of the distinct nature of anomalies—like their scarcity, diversity, undefined boundaries, and the expense involved in gathering vast anomaly data \cite{pang2021deep}. Despite these challenges, researchers have crafted solutions by modifying deep learning techniques to suit anomaly detection, such as using unsupervised learning \cite{fan2018analytical} or feature extraction strategies \cite{himeur2020novel}. These advancements have bridged gaps that were previously insurmountable for standard shallow anomaly detection tools in different areas.

However, as presented in  \cite{pang2021deep}, there are still several challenges remaining that are mostly associated with the data given that anomalies are rare and heterogeneous.  Deep learning algorithms are known to be data-hungry. Hence, supervised deep learning approaches are unpractical and have limited applicability in this subject matter. That is, collecting sizeable labeled data for anomaly detection is mostly difficult and unfeasible. On the other hand, the high false positive rate is another common challenge, especially with unsupervised deep learning-based methods as normal observations are falsely identified as anomalies.   
 Researchers have been attempting to adjust deep learning algorithms to fit the problem and make full use of their capabilities to address this challenge. 
Attempts included applying advanced pre-processing using clustering analysis to identify the diverse types of normal patterns and avoid misidentifying them as anomalies \cite{fenza2019drift}. 

Unsupervised learning methods, which do not use any prior knowledge about anomalies, often underperperform in spotting intricate anomalies. Such elusive anomalies can remain unnoticed for extended periods. Some research, like \cite{tasfi2017deep}, proposes the use of existing historical data. By leveraging both the sparsely labeled data and the vast unlabeled data, they aim to create semi-supervised deep anomaly detection systems. While these methods do not provide a complete solution, they are seen as effective steps toward addressing the issue.
Another often overlooked challenge is how resilient and robust deep anomaly detection methods are in the face of noise and adversarial data. To address this, hybrid techniques that merge deep learning and expert insights have been explored. These solutions enhance the reliability and efficiency of detection but might be more intricate to craft.

\subsection{Contribution of the paper}
When utilizing traditional ML algorithms for anomaly detection, a significant challenge is the high rate of false alarms or false-positive rates. This often occurs because many conventional ML models tend to classify any unfamiliar observation as an anomaly, even if it might be a normal variant not represented in the training set. Theoretically, expanding the training dataset to encompass all possible normal scenarios could mitigate this issue, but practically, this inclusivity is difficult to achieve. Even with an augmented training dataset, a model with a robust generalization capability is essential. In this context, deep learning (DL) models may offer superior generalization compared to traditional ML algorithms.

In line with this understanding, our research introduces, to the best of our knowledge, the inaugural application of a deep feedforward neural network (DFNN) for detecting anomalous energy consumption in buildings. This approach leverages the advanced generalization capabilities of DL to more accurately identify genuine anomalies in energy usage patterns.

Overall, the main contributions of this research lie in advancing the field of anomaly detection in sports facilities through the application of deep learning, providing valuable insights into energy-saving strategies, and presenting a novel approach to improve anomaly detection accuracy and efficiency in real-world settings. These contributions have implications for enhancing users comfort and wellbeing, and sustainability in sports facilities. To summarize, the principal contributions of this article are: 
\begin{itemize}
\item Focusing on anomaly detection in sports facilities, which is crucial for promoting energy-saving practices. By identifying anomalies in energy consumption patterns, the study aims to optimize operational efficiency and reduce unnecessary energy wastage, contributing to sustainability and cost-effectiveness.

\item Exploring the role of machine learning, specifically deep learning techniques, in anomaly detection for sports facilities. By leveraging the power of deep learning models, the study aims to achieve more accurate and efficient anomaly detection compared to conventional approaches.

\item Delving into the challenges and perspectives of applying deep learning methods to anomaly detection in sports facilities. By analyzing the limitations of existing techniques, the study proposes novel solutions and insights to overcome these challenges.

\item Introducing a novel approach for anomaly detection in sports facilities, involving feature extraction from the collected data. This approach is tailored to the specific requirements of sports facilities, considering their unique characteristics and energy consumption patterns.

\item Formulating the anomaly detection problem using DFNNs and incorporating threshold estimation techniques to distinguish anomalies effectively. This contributes to better anomaly detection accuracy and reliability.

\item Proposing methods to reduce false alarms in anomaly detection, ensuring that only genuine anomalies are detected and minimizing false positives. This enhances the overall effectiveness and usability of the anomaly detection system.

\item Empirical Evaluation and Comparative Study: The research conducts comprehensive experiments on real-world datasets, including residential and sport facility data, to evaluate the proposed approach's performance. Through a comparative study, the study demonstrates the superiority of the deep learning-based method over conventional techniques, highlighting the practical implications of the research findings.
\end{itemize}

\subsection{Organization of the paper}


The remainder of the paper is structured as follows. In Section \ref{sec2}, the study delves into the existing literature, discussing both conventional and deep anomaly detection techniques, while also highlighting their drawbacks and limitations. Section \ref{sec3}, presents the core contribution of the paper. It elaborates on the methodology for feature extraction, problem formulation, the deployment of Deep Feedforward Neural Networks (DFNNs), and strategies for threshold estimation and reducing false alarms. The "Experimental Results" in Section \ref{sec4} provide a comprehensive look at the dataset, preprocessing steps, and a discussion of the results. This section also includes a critical comparison with state-of-the-art methods to contextualize the study's findings within the broader research landscape. Finally, Section \ref{sec5} concludes the paper by summarizing the findings, implications, and potential avenues for future research. Each section is meticulously crafted to lead the reader through a logical progression from the existing body of knowledge to the study's novel contributions and findings.

\section{Related Works} \label{sec2}

Anomaly detection of energy consumption can be performed using supervised or unsupervised learning. Supervised learning requires anomaly benchmarks, which are representative of the abnormal energy consumption values against which detected anomalies can be compared to avoid false alarms.

\subsection{Conventional anomaly detection}
In \cite{araya2017ensemble}, the authors propose a pattern-based scheme called collective contextual anomaly detection based on sliding windows (CCAD-SW) for detecting abnormal energy profiles. The method utilizes overlapping sliding windows to identify anomalies. Additionally, they introduce an ensemble anomaly detection (EAD) by combining multiple classifiers through majority voting. The evaluation of this approach is performed using real-world energy consumption data from various buildings in Brampton, Ontario, Canada.
In \cite{liu2017unsupervised}, the authors present an anomaly detection scheme that utilizes a spatio-temporal feature extraction technique based on symbolic dynamics. This technique helps uncover and represent causal interactions among subsystems. The extracted features are then input into a restricted Boltzmann machine (RBM) to learn system-wide patterns and create an energy abnormality detection system.

Yip et al. \cite{yip2018anomaly} introduced a linear programming (LP)-based technique for anomaly detection, aimed at evaluating consumers' energy consumption habits, identifying faulty meters, and preventing potential energy frauds. The method is effective in capturing energy theft attacks on advanced metering infrastructure (AMI), locating sub-meter defects in smart grid (SG) environments, and addressing issues related to non-technical loss (NTL) detection.
Capozzoli et al. \cite{capozzoli2018automated} presented an energy anomaly detection scheme that analyzes energy time-series data from buildings to identify unexpected and unusual power consumption habits. The methodology incorporates an improved symbolic aggregate approximation procedure and optimizes the tuning of time-window length and symbol intervals based on power consumption activities.

Zhang et al. \cite{zhang2021data} proposed an unsupervised anomaly detection approach for integrated energy systems (IESs). Their framework combines four machine learning models, including clustering analysis (CA), knowledge-based (KB), one-class support vector machine (OCSVM), and isolation forest (IF). This approach facilitates the detection and analysis of possible vulnerabilities and threats in an IES, considering modifications and changes in the operation status and run-time of each subsystem.
Do et al. \cite{do2018energy} introduced an unsupervised and scalable abnormality detection scheme using a mixed-variate RBM (Mv-RBM), a principled-probabilistic technique for estimating the density of mixed data. The abnormality score is derived from the free energy extracted from the Mv-RBM, which represents the data's negative log density up to an additive constant. The method extends the detection of abnormalities through multiple levels of data abstraction, leading to the development of the MIXed data Multilevel Anomaly Detection (MIXMAD) solution by constructing an ensemble of mixed DBNs with varying depths.

Rashid et al. propose an unsupervised anomaly detection approach for identifying abnormal daily energy consumption in buildings \cite{rashid2018monitor}. The method utilizes data from smart meters and employs clustering techniques, particularly the local outlier factor (LOF), to group the data into various energy consumption patterns. Local abnormality scores are calculated for each cluster, considering all data points within the cluster to minimize false alarms. Any significant deviation of energy consumption from the prominent pattern identified for a cluster results in flagging it as anomalous.

In Sial et al.'s research, they explore four distinct schemes for detecting abnormal energy consumption in hostel buildings \cite{sial2019detecting}. These schemes include percentage change in consumption (PCC), k-nearest neighbor (kNN), histogram buckets (HB), and principal component analysis (PCA). The meter data is categorized based on factors like the hour of the day, type of the day, and type of power supply. The data is then pre-processed to normalize and fill in missing values, enhancing data analysis. In all four schemes, an anomaly score is calculated, and a predefined threshold using the percentage confidence interval is used to identify abnormal consumption patterns.

Chiosa et al. \cite{chiosa2021data} present a framework for actively detecting abnormal energy consumption at the building level and diagnosing the sub-loads level to identify the source of anomalies. The approach involves using regression trees (RTs) and adaptive symbolic aggregate approximation (aSAX) to identify frequent and infrequent aggregated energy patterns. Association rule mining (ARM) is then employed to uncover the sub-load(s) responsible for the anomalies.
In Xu et al.'s study \cite{xu2021anomaly}, they propose an anomaly detection and dynamic energy performance evaluation method for HVAC systems. They use ARM and clustering analysis to identify energy patterns and relevant association rules for detecting and diagnosing anomalous energy consumption.
Zhou et al. \cite{zhou2021anomaly} focus on the anomaly detection of energy consumption patterns in central air conditioning systems (CACS) by using information entropy (IE) for characterizing daily consumption patterns. This helps mitigate issues with high miss rates and false positives. They also continuously update a dataset of normal daily energy consumption patterns (NDECP) online and flag non-conforming data patterns as anomalous.
Additionally, Bang et al. \cite{bang2019novel} propose a model-based fault detection method using a building simulation model to detect abnormal energy consumption in ventilation systems of buildings. They establish a baseline based on the building simulation model and compare it with actual data using the Chernoff bound method to identify instances of anomalous behavior.







\subsection{Deep anomaly detection}
Deep learning, known as deep anomaly detection (DAD), has gained popularity as a powerful technique for identifying anomalies. In DAD, data inputs receive anomaly scores to detect outliers, or out-of-distribution patterns, with exceptional performance across complex datasets like time series, speech, ECG, images, and videos.

Fan et al. \cite{fan2018analytical} propose an energy consumption anomaly detection method using a deep autoencoder (DAE). They explore various DAE models, including convolutional autoencoders (CAE) and recurrent autoencoders (RAE), forming a DAE-based ensemble approach.
Hollingsworth et al. \cite{hollingsworth2018energy} study energy prediction and machine learning (ML) models to detect energy consumption anomalies. By integrating ARIMA-based energy forecasting with LSTM, they analyze day-to-day operations and remove seasonality and trends from energy observations.
Wang et al. \cite{wang2019power} propose a residual-based approach for energy consumption anomaly detection using LSTM neural networks. The difference between predicted and actual values serves as an indicator of energy consumption status, employing predefined thresholds.
Xu et al. \cite{xu2020hybrid} introduce an unsupervised approach to detect anomalous energy usage in residential buildings using RNN with quantile regression (RNN-QR). They first predict energy consumption and then identify abnormal patterns, showing promise for effective anomaly detection in energy behavior.

Pereira et al. \cite{pereira2018unsupervised} propose a scalable and unsupervised system for anomaly detection in energy consumption time-series data. They enhance the model by incorporating a variational self-attention mechanism (VSAM) into the encoder and decoder, enabling the generation of probabilistic reconstruction scores to effectively identify anomalies.

In another study, Tasfi et al. \cite{tasfi2017deep} introduce a CNN-based approach for detecting consumption anomalies in building automation and management systems (BAMSs). Their supervised CNN-based auto-encoder with branching outputs, the reconstruction, and classification branches, effectively detects and classifies consumption anomalies based on data labels.

Moreover, Fenza et al. \cite{fenza2019drift} employ an LSTM network to forecast consumers' behavior based on energy consumption patterns, distinguishing genuine anomalies from normal behavioral changes. Clustering analysis is used to identify energy consumption profiles, and the LSTM-based model predicts future individual consumption to accurately detect potential anomalies in consumers' behavior.

%

Chahla et al. \cite{chahla2020deep} propose a new framework for daily consumption anomaly detection and isolation. They utilize an NN-based auto-encoder (AE-NN) to effectively detect anomalous consumption patterns and combine it with the K-means clustering algorithm and an LSTM network to localize anomalies throughout the day, considering repeated identical daily consumption patterns due to users' habits.
In another study, Xu et al. \cite{xu2020abnormal} present an anomaly detection framework for ground source heat pump (GSHP) systems in buildings, known for their high energy consumption. They use a mode decomposition-based LSTM algorithm to predict energy consumption and Grubbs' test to identify abnormal system energy consumption based on the difference between predicted and actual values.
Himeur et al. \cite{himeur2020novel} implement a rule-based scheme using the micro-moment paradigm to extract features from power consumption data. This allows them to classify the data into five distinctive classes based on factors like consumption status, occupancy, and appliance use. Their deep neural network (DNN)--based anomaly classifier automatically identifies the five abnormal consumption classes, facilitating efficient detection and categorization of various abnormal behaviors.







\subsection{Drawbacks and limitations}
The existing unsupervised anomaly detection methods face several limitations when applied to big data, including decreased performance and computational efficiency. Statistical methods lack scalability for large-scale data and rely on strict mathematical assumptions that may not hold for real-world high-dimensional data. Although some unsupervised data mining techniques aim to improve efficiency, the associated post-mining workload can be overwhelming, especially in tasks like selecting useful association rules.
Moreover, the performance of existing unsupervised methods heavily depends on the features used, often requiring domain expertise or simple statistics. This lack of data-driven methods for automating feature generation hinders generalization purposes. To overcome these challenges and enhance the applicability of unsupervised anomaly detection in the building field, more advanced methods are needed.
One promising solution to these limitations lies in using autoencoders, which adopt a neural network architecture for unsupervised learning with identical model input and output. The rapid advancements in deep learning have introduced various techniques for analyzing different types of data and training models with advanced architectures, such as deep convolutional autoencoders. Autoencoders enable a data-driven approach to high-level feature extraction, addressing the most challenging aspect of unsupervised anomaly detection, which is feature engineering. Leveraging autoencoders can potentially lead to improved performance in detecting and isolating anomalies in energy consumption data, overcoming the limitations of existing methods.

\begin{center}
\small
\begin{longtable}{
m{0.5cm}
m{2cm}
m{4cm}
m{2cm}
m{6cm}
}
\caption{A summary of anomaly detection of energy consumption  frameworks discussed in this paper}
\label{tab1}\\
\hline 
Work & ML model & Description & Application & Limitation/Advantage \\ \hline
\endfirsthead
\hline
\multicolumn{5}{c}{{Table \thetable\ (Continue)}} \\
\hline
Work & ML model & Description & Application & Limitations/Advantages \\ \hline
\endhead
\hline
\endfoot

\cite{araya2017ensemble} & Autoencoder, SVM & A pattern-based anomaly classifier that detects abnormal energy usage using overlapping sliding windows. & Anomaly detection  & CCAD-SW enhanced the True Positive Rate (TPR) of CCAD by 15\% while decreasing the False Positive Rate (FPR) by 8\%. Additionally, the EAD framework further boosted CCAD-SW's TPR by 3.6\% and lowered its FPR by 2.7\%. \\

\cite{liu2017unsupervised} & SPTN, RBM & A data-driven anomaly detection framework that integrates spatiotemporal feature extraction with anomaly identification. & Anomaly detection & When only a small number of measurements are abnormal for a brief period, no false alarms are detected. Nonetheless, substantial variations among nominal modes can occur in certain instances, and a single-layer RBM might not capture all the features. \\

\cite{yip2018anomaly} & LP & 
An anomaly detection system designed to identify energy theft and malfunctioning meters within smart grids.  & NTL
detection & Unconstrained by the size of consumer power consumption data, yet it necessitates labeled data. \\

\cite{zhang2021data} & CA, KB, OCSVM,IF  & A data-driven method for dynamic analysis of anomalies and vulnerabilities in large-scale integrated energy systems.  & large-scale integrated energy systems &  By gaining a deeper understanding of the system's operating state, its value can be further enhanced, especially considering the high volatility involved in the operation of complex systems. \\

\cite{capozzoli2018automated}& CART, aSAX &Anomaly detection using an improved Symbolic Aggregate approXimation (SAX) process, incorporating optimized tuning of the time window width and symbol intervals to align with the building's energy behavior.& Building energy efficiency& It may lead to information loss, making it challenging to generalize the setting for different types of buildings due to varying occupancy and system operation schedules.\\

\cite{rashid2018monitor} & LOF & Identify unusual daily energy consumption patterns in buildings. & Anomaly detection& Improve the accuracy
of abnormality detection by up to 24\% in the best scenario and
on average by 14\%. However, high FNR for hourly consumption anomalies was reported.\\
\cite{sial2019detecting} & PCC, kNN, HB, PCA & Detect anomalous energy consummation & Anomaly detection & Fails to tackle the issue of potentially high False Positive Rates (FPR).  \\

\cite{chiosa2021data} & RTs, aSAX, ARM  &  Identify abnormal energy consumption at the building level and pinpoint the origin at the sub-loads level. & Anomaly detection and diagnosis & For dynamic deployment, the mean classification accuracy stood at 82.85\%, while for the static deployment, it was recorded at 78.77\%. \\

\cite{xu2021anomaly} & Clustering analysis, ARM & Detect anomalous energy usage of HVAC systems & Commercial buildings & The system is adaptable and modular, allowing for the selection or substitution of components based on specific requirements to forecast the energy consumption of an HVAC system.
\\

\cite{zhou2021anomaly} & IE & Detect anomalous  CACS  daily energy consummation & Anomaly detection  & This online abnormal DECP detection method relies on expert experience for determining characteristic parameters for classification, which may limit its potential application and generalization to different operating conditions. \\

 \cite{bang2019novel} & Model-based & Detect anomalous  energy consummation & Fault detection& The top-down model-based used in this study relies on a simple percentage-based approach for defining threshold limits, which may not be suitable for all building performance contexts and may not account for complex variations in energy consumption.\\

\cite{hollingsworth2018energy} & ARIMA, LSTM & Detect power anomalies by removing seasonality and trend from energy consumption data and comparing it to predictive analysis results, with data provided by Tennessee Valley Authority (TVA). & Residential buildings & Rely on historical data and may not account for unforeseen changes or irregularities in the energy delivery system, potentially leading to inaccuracies in anomaly detection in real-time scenarios. \\

 \cite{wang2019power} &LSTM & Detect anomalous energy consummation & Anomaly detection&  Outperform the ARIMA algorithm by reducing the forecasting error by 22\%, and achieves improved electricity theft identification using actual power consumption data from Chongqing, China. \\

 \cite{tasfi2017deep} &CNN & Detect anomalous  energy consummation & Anomaly detection& Assumes the availability of partially labeled data  \\
 
 \cite{fenza2019drift} & clustering, LSTM  & Predict consumer behavior patterns to differentiate genuine anomalies from regular behavioral variations. & anomaly detection & One inconvenience is the delay in anomaly detection, which stands at a few hours (corresponding to a few tens of observations happening every 15 minutes). This delay may impact real-time responsiveness in detecting and responding to anomalies promptly. \\

 \cite{chahla2020deep} & AE-NN, K-mean clustering, LSTM &  Detects daily anomalous consumption and localize anomalies throughout the day &  Anomaly detection and isolation & It outperforms other tested methods, with higher accuracy, better recall, and a lower false-positive rate. However, the limitations include the potential complexity, reduced interpretability, sensitivity to hyperparameters, limited precision in anomaly detection timing, and the possible lack of generalizability to different datasets and domains.\\

 \cite{xu2020abnormal} & LSTM & Forecast the usage of GSHP pumps and conduct a static analysis of the prediction error to identify unusual consumption patterns.  & Anomaly detection &  The rationality of detected anomalies could only be discussed and verified by field investigation due to limited information available, and the proposed method may not fully account for all factors that can affect energy consumption in the GSHP system.\\

 \cite{himeur2020novel} & DNN  & Using micro-moments to cluster energy usage patterns into various classes, then applying DNN to automatically identify them. & Households and academic buildings&  Achieving 99.58\% accuracy and 97.85\% F1 score were attained under real-world dataset. However, it relies on a rule-based model and may not fully capture complex patterns and variations in energy consumption behaviors. \\

\cite{copiaco2023innovative} &  AlexNet, GoogleNet&  DAD using 2D image representations as features of a supervised deep transfer learning (DTL) approach& Households and academic buildings&   The approach attained F1-scores of 93.63\% and 99.89\% with simulated and real-world energy datasets, respectively. Nonetheless, it depends on supervised learning, necessitating pre-training data labeling for CNN models—a process that may be both time-consuming and labor-intensive. \\

\cite{himeur2022two} &  XGBoost, Isolation Forest  & XGBosst detects energy consumption anomalies of the weekdays while Isolation Forest identifies abnormalities of weekends &  Households &  High accuracy and F1 score while outperforming existing DL-based ADEC methods at a low computational cost. However, it focuses on abnormal energy usage caused by temperature variation and day-offs, which may not cover all possible types of anomalies that could occur in a BIoT environment \\ 
.

\cite{lei2023dynamic}&  PSO& Utilize a dynamic method for detecting building energy consumption anomalies through Particle Swarm Optimization and Unsupervised Anomaly Clustering.  & Unsupervised
anomaly
clustering & The results show that the clustering accuracy of the algorithm can reach more than 80\%, and it can effectively detect the building energy consumption data of two different forms of outliers \\

\cite{lin2023anomaly} &  CNN & Introducing ElectricityTalk, an IoT platform for smart farms, integrating AI mechanisms with farming IoT devices for electric energy prediction and anomaly detection.   &  Smart Farms  & ElectricityTalk identifies all anomalies in actual farm operations, achieving a recall of 1 and a precision exceeding 0.994.\\ 

\hline
\end{longtable}
\end{center}

\section{Proposed Approach}  \label{sec3}
Our proposed method aims to leverage the rich information encapsulated in the energy usage data of sports facilities (aquatic center) to identify patterns and anomalies effectively. By harnessing the power of deep learning, specifically DFNNs, we introduce a comprehensive framework designed to not only detect aberrant energy consumption instances but also minimize false alarms, a common issue in conventional anomaly detection systems. Central to our approach is the sophisticated processing of temporal and contextual data through feature extraction, emphasizing the cyclic nature of time and leveraging occupancy and appliance usage data to create a nuanced understanding of energy consumption patterns.

\subsection{Feature extraction}

To enhance the classifier's ability to learn specific characteristics from observations repeated over time, it is crucial to encode data in a way that (i) facilitates its classification and (ii) improves the quality of machine learning regressors by accurately representing the distance between objects. Data encoding is pivotal for aiding classification. The feature set includes appliance ID, hour of the day, day of the year, occupancy, and power consumption. However, the 'hour of the day' and 'day of the year' exhibit periodicity, meaning that, for example, hour 1 on day 2 follows hour 24 on day 1, and day 1 in the following year comes right after day 365 of the current year. While probabilistic distributions like the von Mises distribution can efficiently model appliance operation times, our goal is to capture the cyclic nature of time and days for use as classification features. Therefore, time and day features are encoded using a sinusoidal approach, educating learners about their periodicity. This cyclic nature is emphasized by transforming each hour and day feature into two components: the cosine and sine, which brings the values of cyclic data closer together. This transformation involves converting both the time incident (\(1 \leq t \leq 24\)) and day stamp (\(1 \leq d \leq 365\)) into their sine and cosine transforms.

The major challenge in anomaly detection methods, particularly in critical fields like cybersecurity, intrusion detection, and medical diagnosis, is the high rate of false alarms (or false positives). This occurs when normal patterns are misclassified as anomalies due to their absence in the training repository. While theoretically expanding the training dataset to encompass all normal scenarios could address this, practically it is unfeasible. A viable solution to mitigate false alarms involves employing deep learning models known for their robust generalization capabilities.

\begin{figure*}[t!]
\centering
\includegraphics[width=1\columnwidth]{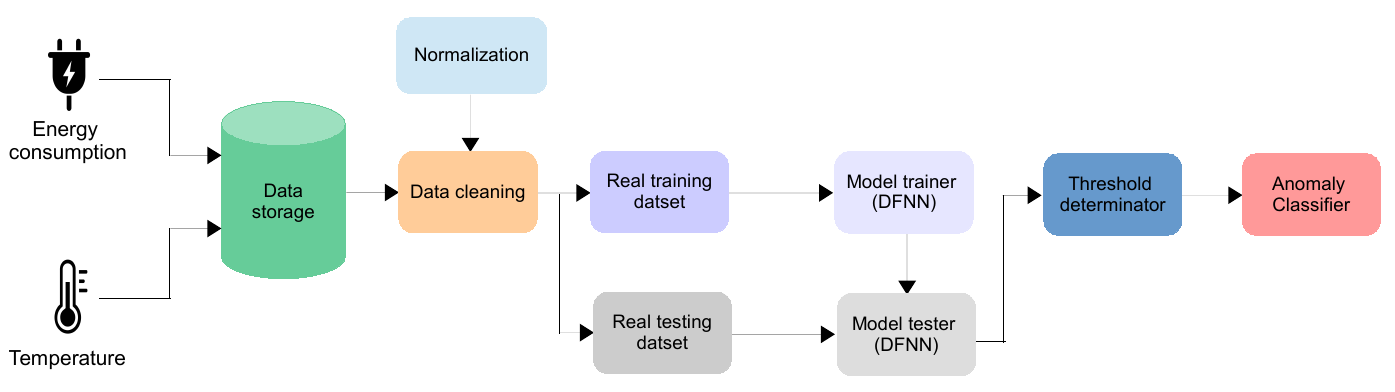}
\caption{Flowchart of the proposed energy consumption anomaly detection approach.}
\label{fig:flowchart} 
\end{figure*}

To this end, with the goal of accurately detecting abnormal energy consumption in buildings, we utilize a substantial dataset for training, coupled with a DFNN model to tackle the overfitting issue prevalent in conventional machine learning classifiers. Figure \ref{fig:flowchart} depicts the flowchart of the proposed anomaly detection scheme.

\subsection{Problem Formulation}
Mathematically speaking, the problem of anomaly detection of energy consumption can be formulated as follows. 
Power consumption observed in a specific building $\widetilde{P}$ is represented s the sum of a function of time $t$ and diverse factors $b_{1}, b_{2},... ,b_{n}$ and random noise $\eta$ as follows:

\begin{equation}
\widetilde{P}=f\left( t,b_{1},~b_{2},~\cdots ,~b_{n}\right) +\eta 
\end{equation}
Our idea to resolve the anomaly detection problem in energy consumption is based on (i) building the aforementioned function, (ii) measuring the noise, and (iii) considering any other additive $\zeta $
 as  anomalous consumption:
 
\begin{equation}
\widetilde{P}=f\left( t,b_{1},~b_{2},~\cdots ,~b_{n}\right) +\eta
+\zeta 
\end{equation}

Aiming to estimate the normal energy consumption, denoted as \(\widehat{P}\), we employ several parameters: (i) \(t\), representing temperature; (ii) \(h\), indicating the time-of-the-day in hours; and labels averaging daily temperature \(\overline{t}\) relative to a predefined threshold. This threshold is instrumental in monitoring the operation of certain devices/systems within the building. Additionally, \(w_{day}\) denotes the label of a working day. Other parameters were individually applied for minor quality enhancements, such as year, month, and day of the week for the site "038", and year, month, day\_of\_week, along with meter values with IDs 875, 896, and 925 for the site "234\_203".

It is worth noting the based on our initial observations, we identified four power consumption $\widehat{P}$ cases with reference to the temperature $\overline{t}$ and $w_{day}$ as follows: \enquote{case 1: $\overline{t} > 17 ^\circ C$ \& $w_{day} = 1$}, \enquote{case 2: $\overline{t} < 17 ^\circ C$ \& $w_{day} = 1$}, \enquote{case 3: $\overline{t} > 17 ^\circ C$ \& $w_{day} = 0$}, and \enquote{case 4: $\overline{t} < 17 ^\circ C$ \& $w_{day} = 0$}.

\subsubsection{DFNNs}
A Multi-Layered Network (MLN) consists of numerous sigmoid neurons and is adept at processing non-linearly separable data. The layers nestled between the input and output are known as hidden layers, which manage the intricate non-linear relationships between the input and output.

\textbf{step 1:} To model energy consumption $\widehat{P}$ in each building as a function of the diverse parameters already discussed in previous section, we use a DFNN. The DFNN model is utilized using five hidden layers, where 1024 neurons are used in each layer, and by adopting a ReLU activation. Moving on, a probability of dropout of 0.5 is employed with the dropout layer that has been embedded after every hidden layer. Moreover, as it is already considered that there are some anomaly observations in the energy consumption footprints, the mean absolute error (MAE) is employed as a loss function. Additionally, to optimize the model, we deploy the Adam optimizer with 5-10 epochs. In this context, after completing the training, $\widehat{P}$ is estimated using this model for the same points. High dropout was used to prevent overfitting

\textbf{step 2:} After estimating $\widehat{P}$, we calculate the difference with the observed power consumption $\widetilde{P}$:

\begin{equation}
\Delta =\widehat{P}-\widetilde{P}
\end{equation}

\textbf{Ste3:} We normalize this difference to make it a metric in some sense:

\begin{equation}
\varepsilon =\left\vert \frac{\Delta -\overline{\Delta }}{\sigma \Delta }%
\right\vert 
\end{equation}

where $\overline{\Delta }$\ refers to the mean estimation of $\Delta $ by
performing arithmetic average, while $\sigma \Delta $ that is the estimation of the std for $\Delta $, which is estimated as follows:

\begin{equation}
\sigma \Delta =\frac{1}{n-1}\sum \left( \Delta -\overline{\Delta }\right)
^{2}
\end{equation}%

If the number of abnormal points in the dataset is known and fixed, thresholds $Thr$ for $\varepsilon$  can be
estimated to get the desired number of abnormal points. All points with $\varepsilon > Thr$ are marked as
anomalies. Because the number of abnormal points is unknown in the dataset and there may be
abnormal points in another sense, different thresholds were used to get high scores on the leaderboard (LB).

\subsection{Threshold estimation}
Since unsupervised anomaly detection is employed, both the number and types of energy consumption anomalies remain undefined, potentially including various forms of anomalies. Consequently, a range of threshold values is used to optimize anomaly detection performance. An initial threshold is set at 3, corresponding to the 99.7

\begin{figure*}[t!]
\centering
\includegraphics[width=0.6\columnwidth]{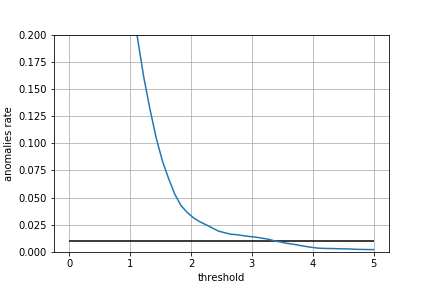}
\caption{Example of estimating the adequate threshold for an anomaly rate of 1\% (black line), i.e. the anomaly rate vs. anomaly detection threshold. }
\label{fig2} 
\end{figure*}

Thresholds ranging from 2.5 to 4.5 were determined for various buildings and parameter areas. A graph similar to that in Fig. \ref{fig2} assisted in selecting appropriate threshold values to achieve the targeted rate of abnormal points for each of the models depicted in Fig. \ref{fig4} for each building. Fig. \ref{fig4} also presents a scatter plot illustrating the correlation between observed and estimated normal energy consumption.


\subsection{False alarms reduction}
We define an instance as an outlier if the prediction error exceeds a predefined threshold. To avoid over-fitting, we split the data into training and testing sets to ensure each instance occurs only once in the train set and test set, as Fig. \ref{fig5} shows.

\begin{figure*}[t!]
\centering
\includegraphics[width=1\columnwidth]{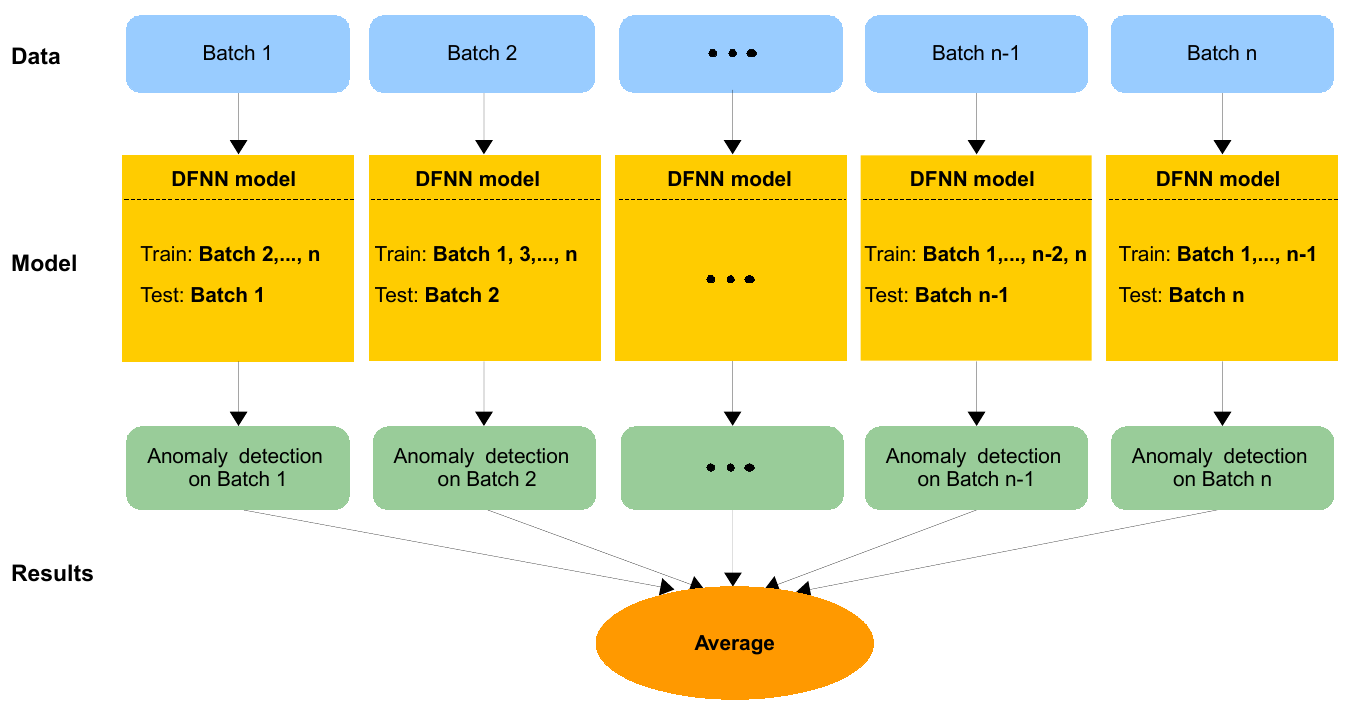}
\caption{Proposed solution to reduce false alarms.}
\label{fig3} 
\end{figure*}

\section{Experimental Results} \label{sec4}
Aiming at evaluating the performance of the proposed anomaly detection scheme, it is important to firstly assess the energy prediction scheme used, which is part of the anomaly detection framework. In this context, different criterions have been deployed, among them the MAPE, MAE and RMSE are defined as follows:

\begin{equation}
MAPE=\frac{1}{n}\sum_{i=1}^{n}\frac{\widehat{X_{i}}-X_{i}}{X_{i}}\times 100
\end{equation}

\begin{equation}
RMSE=\sqrt{\frac{1}{n}\sum\limits_{i=1}^{n}\left[ \widehat{X_{i}}-X_{i}%
\right] ^{2}}
\end{equation}

\begin{equation}
MAE=\frac{1}{n}\sum\limits_{i=1}^{n}\left\vert \widehat{X_{i}}%
-X_{i}\right\vert 
\end{equation}
where , $\widehat{X_{i}}$ refers to the predicted value; $X_{i}$ is the observed value; and n represents the sample size.
Indeed, MAPE is adequate to evaluate the results of any energy forecasting model due to its relative values. Typically, as the MAPE is not impacted by the unit or the length of the observed and predicted energy consumption time-series, it can well indicate their relative difference. In this context, a MAPE rate of 10\% of lower should indicate a high prediction accuracy. While MAPE rates of 10–20\%, 20–50\%, and over 50\% refer to good, reasonable, and inaccurate prediction accuracies, respectively.
Moving on, the RMSE measures the square error of the predicted energy consumption time-series in comparison to the observed data before calculating the square root of the summation value.
Due to that fact that the RMSE measures the average of the squared the errors, the values having large errors are weighted heavily and therefore the unacceptably large
differences are revealed. On the contrary, MAE estimates the average amplitude of errors observed and predicted energy consumption time-series without disregarding the direction of errors. Thus, MAE could calculate the continuous variables because all the individual differences are weighted equally.

The validation of the proposed method has been conducted in two scenarios. First, an existing dataset was used to predict the energy consumption and then detect anomalous energy usage.
Second, energy consumption of the aquatic center at Qatar University was recorded, energy consumption was forecasted, and abnormal energy consumption

\subsection{Dataset description}
Although the experiment was situated at the aquatic center of Qatar University, it unfolded in an authentic real-life environment, closely mimicking typical operational conditions. This setting provided a realistic backdrop where variations in power consumption could naturally emerge, reflecting the genuine fluctuations that occur in day-to-day operations. These variations were largely attributed to the activities of individuals within the offices, encompassing a range of behaviors from the use of electrical appliances to lighting and heating adjustments, each contributing to the dynamic power profile of the setting.

The experiment was further bolstered by the use of commercially available smart meters, a choice that reflects the technology's widespread applicability in real-world scenarios. These smart meters were intricately connected to a server operated by the manufacturer, ensuring a seamless flow of data and allowing for real-time monitoring and analysis. This connection not only facilitated the accurate recording of power consumption data but also enabled the exploration of how such devices can be integrated and utilized in broader energy management systems. The integration with the manufacturer's server is indicative of the potential for remote monitoring and diagnostics, a feature increasingly vital in modern energy management solutions.

\begin{figure*}[t!]
\centering
\includegraphics[width=1\columnwidth]{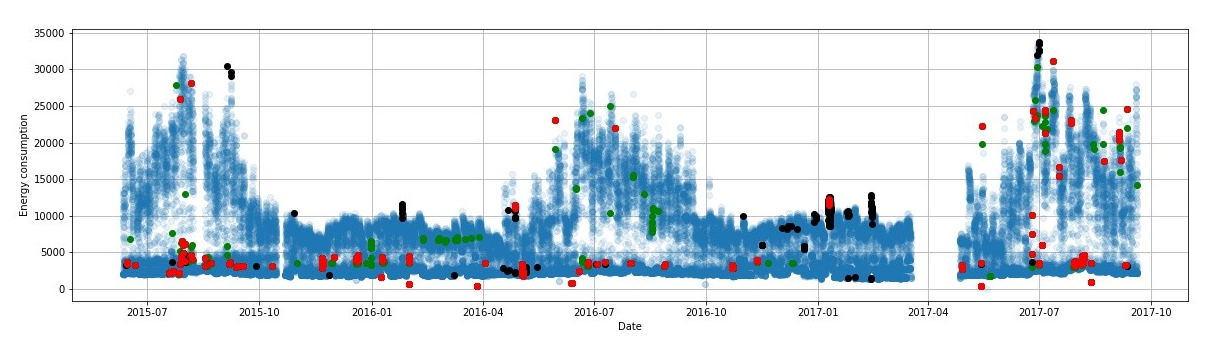}
\caption{Energy consumption anomalies of building 1, where. (i) green points refer to abnormalities detected using DFNN, (ii) black points are abnormalities identified using K-Nearest Neighbors (KNN), and (iii) red points represent final labeling.}
\label{fig4} 
\end{figure*}



\subsection{Preprocessing}
First of all the data was reformatted as a table, including the timestamp, measurement 1, measurement 2, temperature, competition, training, or off-season patterns, etc. Moving on, data preprocessing techniques, including  normalization, scaling, and handling missing values have been deployed to ensure the effectiveness of anomaly detection algorithms.

\subsection{Result discussion}

\begin{figure*}[t!]
\centering
\includegraphics[width=0.7\columnwidth]{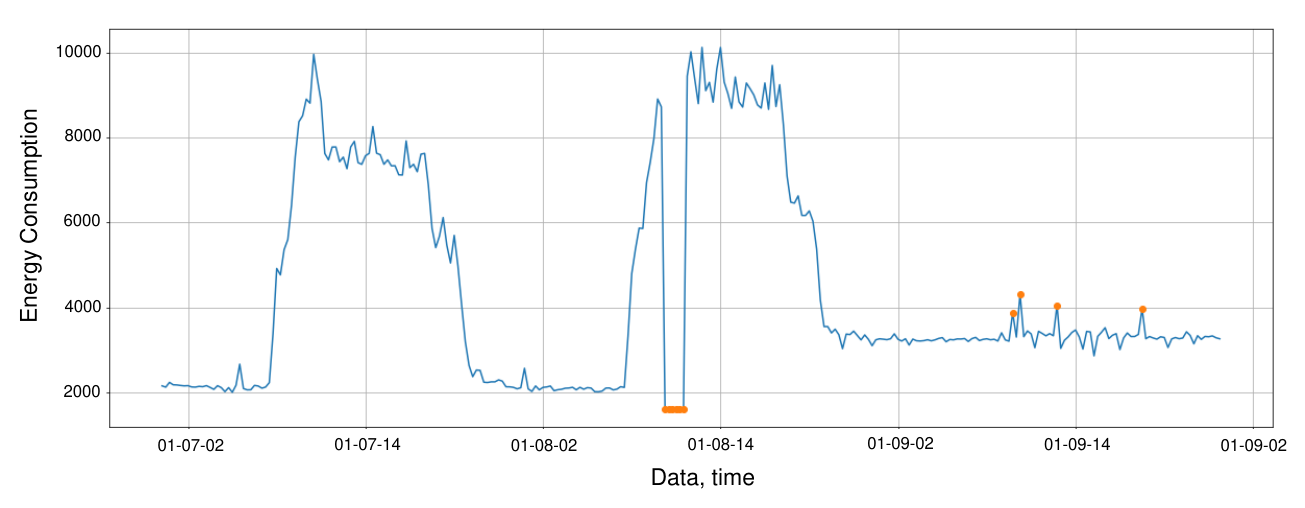}\\
(a) \\
\includegraphics[width=0.7\columnwidth]{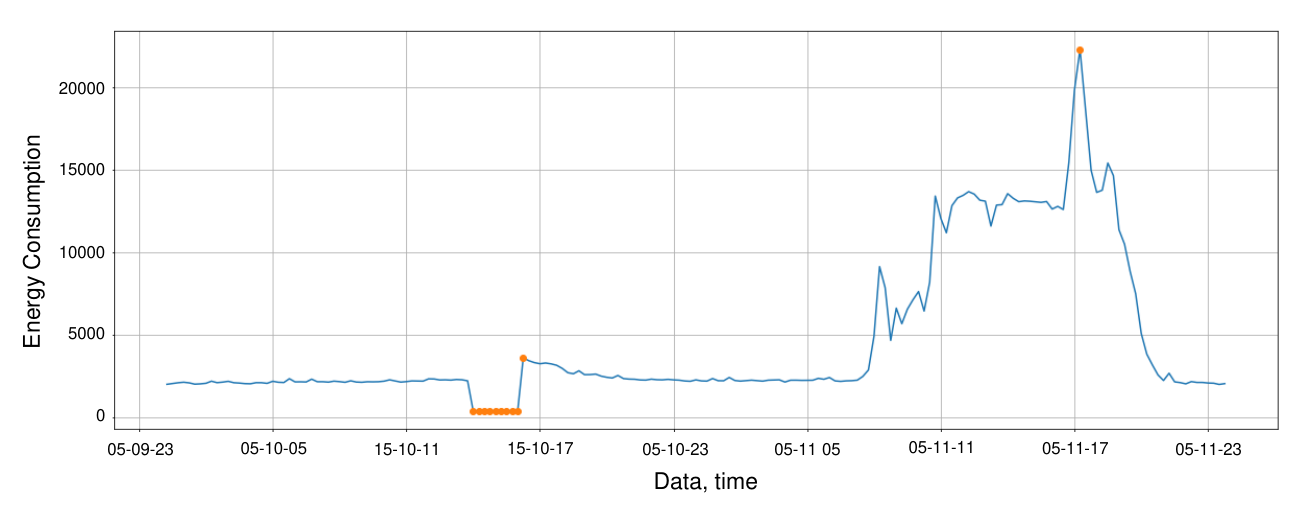}\\
(b) \\
\includegraphics[width=0.7\columnwidth]{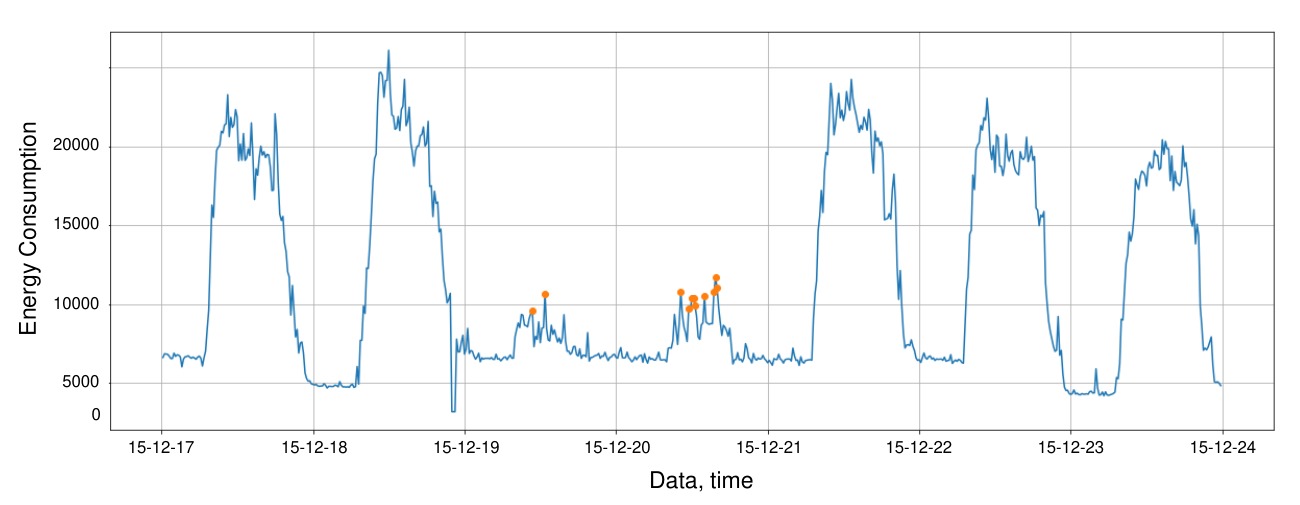}\\ 
(c) \\
\includegraphics[width=0.7\columnwidth]{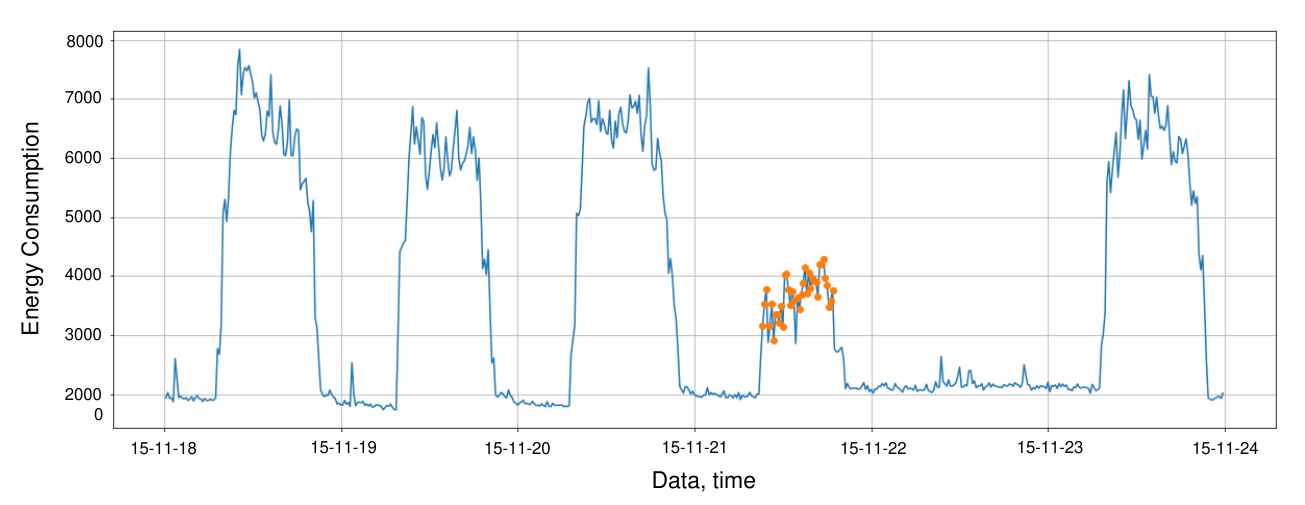} \\
(d)\\
\caption{Types of anomalies found with the method: (a) extremely low consumption and several smaller outliers, (b) low consumption and single abnormal point, (c) high energy consumption for the day, and (d) huge number of outliers for the day.}
\label{fig5} 
\end{figure*}

The anomaly detection method applied to the energy consumption data has revealed various types of anomalies, each providing valuable insights into the building's energy systems and potential areas of concern. Below are the key types of anomalies that were detected:

\begin{itemize}
\item Extremely High/Low Energy Consumption Over a Short Period of Time: One prominent type of anomaly identified by the method is instances of unusually high or low energy consumption occurring within a short timeframe. These anomalies are represented by sharp spikes or dips in the energy consumption graph. Such occurrences can be indicative of abnormal system behavior within the building and may warrant further investigation. For example, sudden spikes in energy consumption could point to malfunctioning equipment, leakages, or inefficiencies in the building's systems. On the other hand, sudden drops in energy consumption might indicate the temporary shutdown of certain systems or the presence of energy-saving measures. Detecting such anomalies can be crucial in identifying potential issues and allocating resources more efficiently during scheduled inspections or maintenance checks, as portrayed in Figs. \ref{fig5}(a) and (b).

\item High/low energy capacity for the time of day: The anomaly detection method also revealed anomalies related to high or low energy capacity occurring at specific times of the day. These anomalies are characterized by unusual fluctuations in energy consumption patterns that deviate significantly from the expected behavior. One prominent example is the detection of abnormal energy usage in the central conditioning system during certain hours. Fig \ref{fig5}(c) illustrates instances where the energy capacity for the time of day shows irregular spikes or drops, indicating potential malfunctions or inefficiencies in the central conditioning system. Identifying such anomalies can be crucial for building managers and maintenance teams as it enables them to promptly address issues with critical equipment. Malfunctions in the central conditioning system can lead to uncomfortable indoor conditions for occupants and result in increased energy wastage. By detecting these anomalies, building operators can take immediate corrective actions, such as conducting maintenance checks or repairs, to ensure the smooth and efficient operation of the central conditioning system.

\item An additional type of anomaly detected by the method is the extremely high or low energy capacity observed for specific days of the week. Figs. \ref{fig5}(a), (b), and (d) showcase instances of abnormally low energy consumption on certain days. These patterns can serve as critical indicators of potential malfunctions in the energy systems of the building. When significant deviations in energy capacity occur consistently on specific days of the week, it could signal the presence of underlying issues affecting the building's energy efficiency. For example, abnormally low energy consumption on certain days may suggest that certain systems or equipment are not operating optimally, leading to potential energy leaks or inefficiencies. Conversely, extremely high energy capacity on certain days might indicate excessive energy usage, which could be attributed to malfunctioning equipment or poor energy management practices. By recognizing these anomalies, building managers and maintenance teams can promptly investigate and diagnose the root causes of these irregular patterns. Addressing critical malfunctions in the energy systems at an early stage can prevent further energy wastage and reduce the risk of costly breakdowns. Moreover, it allows building operators to implement targeted measures to optimize energy consumption on specific days, thus enhancing the building's overall energy performance.

\end{itemize}

\subsection{Comparison with the state-of-the-art.}
This section provides a comparative analysis of various machine learning (ML) models and their effectiveness in anomaly detection, as measured by accuracy and F1-score metrics. Typically, Table \ref{tab2} presents a comparison between recent ADEC frameworks.
In this regard, \cite{himeur2022two} combines XGBoost and Isolation Forest (IF), yielding an accuracy of 83.11\% and an F1-score of 80.73\%. This dual-model strategy is particularly adept at handling the different patterns of weekdays and weekends, including holidays, which is a smart way to address the varying nature of anomalies in energy consumption.

Pena et al.'s rule-based algorithm achieves slightly lower accuracy and F1-score of 82.45\% and 80.25\%, respectively. Their hybrid approach, which merges prediction-based and rule-based methods, offers a nuanced way to identify abnormal energy consumption by considering the 'surprise' factor. This method's strength lies in its ability to discern between genuine anomalies and mere predictive errors, which is crucial for minimizing false alarms in anomaly detection.

Another approach proposed in \cite{himeur2021smart} involves a hand-picking method combined with an LB probing algorithm, resulting in an accuracy of 81.44\% and an F1-score of 80.67\%. This method focuses on a meticulous examination of anomalies, categorizing observations by similar time-related features, and probing each meter individually. Although this method is more labor-intensive and time-consuming, it provides deep insights into the nature of anomalies and contributes to a more nuanced understanding of the problem.

The proposed method in the study employs DFNN and achieves the highest accuracy and F1-score among the compared models, at 92.33\% and 89.91\% respectively. The strength of this method lies in its use of advanced neural network models which are adept at capturing complex patterns and relationships in data. The discussion on potential improvements with additional data suggests that this model is not only robust but also adaptable and scalable.
Overall, while all methods provide valuable approaches to anomaly detection, the proposed DFNN model shows the most promise with its high accuracy and adaptability. However, each method has its unique advantages and could be preferred in different scenarios depending on the specific requirements and constraints of the task at hand. Future research might focus on combining the strengths of these various approaches to develop even more sophisticated and accurate methods for anomaly detection.

\begin{table}[t!]
\caption{Results comparison with the state-of-the-art.}
\label{table:1}
\label{tab2}
\small
\begin{tabular}{
m{22mm}
m{20mm}
m{10mm}
m{10mm}
m{80mm}
}
\hline
\textbf{Place} & \textbf{ML models} & \textbf{Accuracy} & \textbf{F1-score} & \textbf{Method's description}                                                                                                                                            \\ \hline
Himeur et al. \cite{himeur2022two}              & XGBoost + IF            & 83.11               & 80.73                & Two models have been combined to identify anomalies: XGBoost for weekdays and Isolation Forest (IF) for weekends (including holidays).                                       \\ \hline
Pena et al. \cite{pena2016rule}              & Rule-based algorithm               & 82.45               & 80.25                & A hybrid approach, blending prediction-based and rule-based methods, is utilized to identify abnormal energy consumption. The core concept involves training a machine learning model to forecast future energy usage. The 'surprise' factor of the model is assessed by examining the discrepancy between predicted and actual consumption. A significant deviation indicates either a modeling error or an instance of abnormal energy usage. In instances of detected overconsumption, the aim is to distinguish genuine anomalies from mere inaccuracies in the model's predictions.\\ \hline
Himeur et al. \cite{himeur2021smart} & Hand-picking method + LB probing algorithm                 & 81.44               & 80.67        & A hand-picking method and an LB probing algorithm were employed, leveraging an optimization metric that imposes a significant penalty for false positives. This approach led to the identification of concrete examples of anomalies. The aim was also to comprehend the nature of anomalies. Observations with similar time-related features were categorized, and 4 sigma deviations were identified as anomalies. Subsequently, these groups were manually examined, and LB probing was conducted for each meter\_id individually. After about a week, the anticipated score was achieved. Thus, the strategies for identifying initial anomaly candidates may prove beneficial.\\ \hline
Proposed    & DFNN        & 94.33              & 92.91                & Models were created for the identification of anomalies using k-nearest neighbors and DFNN models. Subsequently, the types of anomalies detected by these methods were explored, and the potential for improving anomaly identification with the incorporation of additional data was examined. \\ \hline
\end{tabular}
\end{table}

\section{Conclusion} \label{sec5}
In conclusion, the possibility of improving anomaly detection in energy consumption lies in several key areas. First and foremost, obtaining more precise information about competition, training, and off-season patterns can lead to a significant enhancement in the quality of the models. By incorporating accurate and detailed holiday data, the models can better account for the impact of holidays on energy consumption patterns, allowing for more accurate anomaly detection during these periods.
Secondly, incorporating information about equipment operation modes, such as the temperature thresholds for turning central conditioning or heating systems on and off, can further improve the model's ability to detect anomalies related to specific operational states of the building.

Moreover, exploring the specific usage and goals of the anomaly detection system can be valuable for feature engineering. Understanding which types of abnormal behaviors are truly critical for the particular needs of the application can lead to the selection of relevant features and improve the overall performance of the models.
Additionally, introducing an additional model capable of capturing both low and high-frequency processes as long-term and short-term non-stationarities can be beneficial. This approach allows for a more comprehensive analysis of anomalies at different time scales, enabling the detection of anomalies that occur over extended periods or exhibit rapid fluctuations.

Lastly, approaching the problem as a time-series task and focusing on searching for outliers based on previously seen data only can lead to more accurate and targeted anomaly detection. By marking only the previously unseen data points as anomalies, the models can better adapt to changes in energy consumption patterns over time and avoid false alarms based on historical data.
Incorporating these improvements and utilizing advanced techniques in anomaly detection, such as more sophisticated machine learning algorithms and ensemble methods, holds the potential to significantly enhance the accuracy, efficiency, and real-world applicability of energy consumption anomaly detection systems. These improvements can play a crucial role in achieving higher energy efficiency, reducing costs, and ensuring users comfort and wellbeing.



\end{document}